\documentclass{ws-ijmpa}

\begin{document}

\markboth{Liang Zuo-tang}
{Spin structure of nucleon and
 spin transfer in high energy fragmentation process}

\catchline{}{}{}{}{}
%
%%%%%%%%%%%%%%%%%%%%%%%%%%%%%%%%%%%%%%%%%%%%%%%%%%%%%%%%%%%%%%%%%%%%

\title{Spin structure of nucleon and \\ 
spin transfer in high energy fragmentation process}
\author{\footnotesize Liang Zuo-tang}
\address{Department of Physics, Shandong University, Jinan, Shandong 250100, CHINA}
\maketitle

\pub{Received (Day Month Year)}{Revised (Day Month Year)}

\begin{abstract}
Spin transfer in high energy fragmentation process is determined 
by the hadronization mechanism and spin structure of hadrons.
It can be studied by measuring the polarizations of hyperons 
and/or vector mesons in $e^+e^-$ annihilation, 
in the current fragmentation region of polarized deeply inelastic 
$lN$-scatterings, 
and high $p_T$-jets in polarized $pp$-collisions. 
Theoretical calculations have been made using different models.
In this talk, I will briefly summarize the main features of the models, 
the results obtained and the comparison with available data.
They can be used for future tests by experiments.

\end{abstract}

\section{Introduction}	%) A SECTION HEADING
In this talk, I would like to use this opportunity to stress one 
problem that we met in studying the polarization of hadrons produced 
in high energy reactions. 

It is now well-known that, 
for the spin compositions of nucleon, 
we have two kinds of results: 
One is obtained from the SU(6) wave-function, 
the other from the polarized deeply inelastic  
$lN$ scattering data and other related knowledges.
We denote them by SU(6) and DIS pictures in the following. 
I would like to emphasize that both of them can be extended 
to other baryons in the same flavor SU(3) octet as the nucleon.[1] 
This is because in obtaining the DIS results for nucleon, 
we have already used the flavor SU(3) symmetry. 
The results can be found in different literature, 
e.g., table 1 of [2], which I include also here (table 1).
We see that these results are very much different 
from each other.  
We have therefore a very practical question: 
Which should we use when we calculate the  
polarization of hadrons produced in high energy reactions 
from the polarizations of quarks? 
There exist two classes of theoretical treatments:
One of them simply uses SU(6), the other uses DIS. 
But there is no discussion about the question which 
should be used before [2]. 

In [2], we first explicitly pointed out 
the problem and showed that $\Lambda$ in $e^+e^-$
annihilation at the $Z^0$ pole is an ideal place to study it.  
We made the calculations on $\Lambda$ polarization  
using the DIS picture,  
compared the results with those from SU(6) obtained by 
Gustafson and H\"akkinen earlier [3], and the 
available data from ALEPH [4] and later [5] also 
those from OPAL [6] (see Fig.1). 
We found that the data can unfortunately not be able 
to distinguish between the two pictures although  
it seems that the SU(6) results fit the data better. 
We therefore made a systematical study [2,5,7] for different 
cases in this connection. 
In the following, I will briefly summary 
the results obtained. They can be used for further 
tests by future experiments.
Similar calculations have also been carried out by 
other groups [e.g. 8-11].
\\[-0.6cm]
  
\begin{table}
\tbl{Sin compositions of quarks in the octet baryons.
(See [2] for details.)}
%{\begin{tabular}{@{}cccc@{}} \toprule
{\begin{tabular}{l|clcc|clcc}%{l||c|c|c||c|c|c} %\toprule
\hline 
& & & $SU(6)$ & DIS & & &$SU(6)$ & DIS  \\ \hline
%%%%%%%%%%%%%%%%%%%%%%%% proton and neutron %%%%%%%%%%%%%%%%%%%%%%
$\Delta U$ &   & $(\Sigma+D)/3+F$ & 4/3  & 0.79 
           &   & $(\Sigma-2D)/3$  & -1/3 & -0.47 \\ %\hline  
$\Delta D$ &$p$& $(\Sigma-2D)/3$  & -1/3 & -0.47
           &$n$& $(\Sigma+D)/3+F$ & 4/3  & 0.79  \\ %\hline
$\Delta S$ &   & $(\Sigma+D)/3-F$ & 0    & -0.12 
           &   & $(\Sigma+D)/3-F$ & 0    & -0.12 \\ \hline %\hline 
%%%%%%%%%%%%%% Lambda and Sigma %%%%%%%%%%%%%%%%%%%%%%%%
$\Delta U$ &   & $(\Sigma-D)/3$  & 0    & -0.17 
           &   & $(\Sigma+D)/3$  & 2/3  & 0.36  \\ %\hline  
$\Delta D$ &$\Lambda$  & $(\Sigma-D)/3$  & 0    & -0.17 
           &$\Sigma^0$ & $(\Sigma+D)/3$  & 2/3  & 0.36  \\ %\hline
$\Delta S$ &   & $(\Sigma+2D)/3$ & 1    & 0.62  
           &   & $(\Sigma-2D)/3$ & -1/3 & -0.44 \\ \hline %\hline 
%%%%%%%%%%%%%% Sigma %%%%%%%%%%%%%%%%%%%%%%%%
$\Delta U$ &   & $(\Sigma+D)/3+F$  & 4/3    & 0.82 
           &   & $(\Sigma+D)/3-F$  & 0  & -0.01  \\ %\hline
$\Delta D$ &$\Sigma^+$ & $(\Sigma+D)/3-F$  & 0    & -0.10 
           &$\Sigma^-$ & $(\Sigma+D)/3+F$  & 4/3  & 0.82  \\ %\hline
$\Delta S$ &   & $(\Sigma-2D)/3$ & -1/3    & -0.44  
           &   & $(\Sigma-2D)/3$ & -1/3 & -0.44 \\ \hline %\hline
%%%%%%%%%%%%%%%%%%%%% Xi %%%%%%%%%%%%%%%%%%%%%%%%% 
$\Delta U$ &   & $(\Sigma-2D)/3$  & -1/3 & -0.44 
           &   & $(\Sigma+D)/3-F$ & 0    & -0.10 \\ %\hline  
$\Delta D$ &$\Xi^0$  & $(\Sigma+D)/3-F$ & 0    & -0.10 
           &$\Xi^-$  & $(\Sigma-2D)/3$  & -1/3 & -0.44 \\ %\hline
$\Delta S$ &   & $(\Sigma+D)/3+F$ & 4/3  &  0.82 
           &   & $(\Sigma+D)/3+F$ & 4/3  &  0.82 \\ \hline 
\end{tabular} }
\end{table}

\begin{figure}
\centerline{\psfig{file=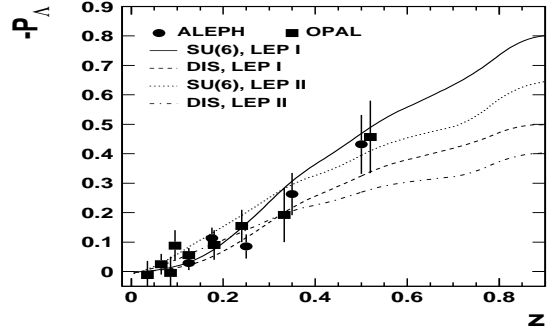,width=9cm,height=4cm}}
\vspace*{8pt}
\caption{$P_\Lambda$ in $e^+e^-\to\Lambda X$.
The data are for LEP I [4,6]. The figure is taken from [5].} 
\end{figure}

\section{The calculation method}
To test different pictures for spin transfer in 
the fragmentation process $q(pol)\to h+X$, we need:
(1) to produce a $q$ beam with known polarization,
(2) to measure the polarization of $h$.
Hence, hyperon polarizations ($P_H$) in the following 
three cases are best suitable:
(a) $e^+e^-\to Z^0\to HX$;
(b) current fragmentation region in polarized $lN\to l'HX$;
(c) high $p_T$ jets in polarized $pp$ collisions. 
This is because, here we can 
(i) separate fragmentation from the others, 
(ii) calculate the polarization of the quark before fragmentation, 
(iii) measure the polarizations of hyperons easily.

To calculate $P_H$ in $q(pol)\to H+X$, 
we divide the produced $H$ into the following four groups:
(A) directly produced and contain the initial $q$;
(B) directly produced but do not contain the initial $q$;
(C) decay contribution from heavier hyperons $H_j$ that are polarized;
(D) decay contribution from $H_j$ that are unpolarized.
That is, 
\begin{equation}
D_q^H(z)=D_q^{H(A)}(z)+D_q^{H(B)}(z)+D_q^{H(C)}(z)+D_q^{H(D)}(z),
\end{equation}
where $D_q^H(z)$ is the fragmentation function.
Similarly, for polarized case, we denote, 
$\Delta D_q^H(z)\equiv D_q^H(z,+)-D_q^H(z,-)$, 
(the $+$ or $-$ means that the produced 
$H$ is polarized in the same or opposite direction as the initial $q$),
and we have,
\begin{equation}
\Delta D_q^H(z)=\Delta D_q^{H(A)}(z)+\Delta D_q^{H(B)}(z)+
\Delta D_q^{H(C)}(z)+\Delta D_q^{H(D)}(z).
\end{equation}
Clearly, there is no contribution for group (D) to $\Delta D$, 
and it is assumed [2,3] that there is no contribution from group (B) either. 
Hence, we have, 
\begin{eqnarray}
&\Delta D_q^{H(A)}(z)=t^F_{H,q} D_q^{H(A)}(z),\ \ \ \ \ \ \ \ \ \ 
&\Delta D_q^{H(B)}(z)=0,\nonumber \\
&\Delta D_q^{H(C)}(z)=\sum_jt^D_{H,H_j} \Delta D_q^{H_j}(z),\ \ \ \ 
&\Delta D_q^{H(D)}(z)=0.
\end{eqnarray}
Here $t^F_{H,q}$ is the fragmentation spin transfer factor 
and is taken as $t^F_{H,q}=\Delta Q^H/n_q$,
where $\Delta Q^H$ is the fractional contribution of 
spin of quark of flavor $q$ to $H$ as given in table 1, 
$n_q$ is the number of valence quarks of flavor $q$ in $H$.
Clearly $\Delta Q^H$ is different in SU(6) or DIS picture. 
This is the place where different pictures come in.
$t^D_{H,H_j}$ is the decay spin transfer factor 
in $H_j\to H+X$. It is determined by the decay 
and is independent of the pictures for spin 
transfer in fragmentation. 

Here, I would like to emphasize the following two points.

First, the classification of $H$ into the above-mentioned 
four groups is independent of the polarization.
All the $D_q^{H(\alpha)}$'s %($\alpha=A,B,C,$ and $D$) 
can be calculated using our knowledge on 
hadronization in unpolarized case.
In fact, in the Feynman-Field-type of cascade fragmentation models,
(A) is just the first rank hadron and the 
$z$-distribution $D_q^{H(A)}(z)$, 
usually denoted by  $f_q^H(z)$ 
is a basic input of the model.  
Practically, all of them can be easily calculated using a 
Monte-Carlo event generator. 
The results are quite stable. 
Hence, the $z$-dependence of 
$P_H$ in this model is empirically known without 
any input in connection with  polarization effects. 
This is a good point to test the model. 
In Fig.1, we see that although different pictures 
lead to different $P_\Lambda$ but the $z$-dependence 
is essentially the same, and it is also in agreement 
with the data. 
This is a strong support of the calculation 
procedure presented above. 

Second, for decay contribution, 
presently, we have data on $t^D_{H,H_j}$ for 
$J^P=(1/2)^+$ octet $H_j$'s.
Hence there is little uncertainty here. 
But for decuplet hyperons, there is neither data 
for $t^D_{H,H_j}$ nor $\Delta Q^H$ in DIS picture.
We can only make a rough estimation by invoking 
the simple quark model for both of them.
There is a strong model dependence, especially
for $\Delta Q^H$, the estimation is definitely too crude.
Hence, to make a good test of different pictures for 
spin transfer in fragmentation, 
we should choose the places to avoid the contribution from 
decuplet hyperon decay.\\[-0.6cm]

\section{Results and discussions}
\underline{(I) $e^+e^-\to HX$}: 
Results for different hyperons have been obtained [2,5]. 
We would like to have flavor separation 
in particular for $u$ or $d$ to $\Lambda$. 
We found that it is impossible to do it 
in $e^+e^-$ annihilation at $Z$-pole 
where $s\to\Lambda+X$ dominates.

\underline{(II) Polarized $e^-N\to e^-HX$ or $\nu_\mu N\to\mu^-HX$:}  
Here, we have almost automatic flavor separation 
since $u\to H+X$ dominates the current fragmentation regions, 
and we have the advantage to study 
both longitudinally and transversely polarized cases 
in $e^-N$ collisions.
I would like to emphasize the following two points 
showed by the results: 
(1) There is a quite large contribution from 
heavier hyperon decay to $\Lambda$ in these reactions 
and the final result for $P_\Lambda$ is small in most cases. 
In particular, in $\nu_\mu N\to\mu^-HX$, we have a 
significant contribution from $\Lambda_c\to\Lambda X$, 
the spin transfer from which is completely unknown.  
Hence it is {\it not} a good choice to use $\Lambda$ 
in these reactions to test different pictures.
In contrast, there is almost no contribution from 
heavier hyperon decay to $\Sigma^+$ and $P_{\Sigma^+}$ is large. 
(2) In the energy region of the presently 
available experiments such as HERMES[12] and NOMAD[13], 
it is impossible to separate the struck quark fragmentation 
from the target remnant contribution. 
In fact, the target remnant contributions 
dominate even in the middle of the so-called 
current fragmentation region in this case [7]. 
It is thus difficult to use these data 
to test the different pictures. 
One has to go to higher energies.

\underline{(III) Polarized $pp\to HX$ at high $p_T$}: 
Here we reached similar conclusions as in (II), i.e., 
the contributions from heavier hyperon decay to $\Lambda$ 
is high and it is more suitable to use 
$p(pol)p\to \Sigma^+X$ to test different pictures. 
For details, see [7].

Besides, we found that spin alignments of vector 
mesons can also provide useful information in this connection. 
Data from LEP exits, 
calculations have been made and compared with them. 
Further predictions for deeply inelastic $lN$ scattering 
or $pp$ collisions have been made.  
Interested readers are referred 
to [14] and the references cited there.

I thank the organizers for inviting me to present this invited 
contribution to the conference. 
This work was supported by NSFC under No. 10175037.

\end{document}